# Synthesis and Thin Films of Thermally Robust Quartet ($S$ = 3/2) Ground State Triradical

Chan Shu[†], Maren Pink[‡], Tobias Junghoefer[§], Elke Nadler[§], Suchada Rajca[†], Maria Benedetta Casu*[§] and Andrzej Rajca*[†]

[†]Department of Chemistry, University of Nebraska, Lincoln, Nebraska 68588-0304, USA.
[‡]Department of Chemistry, Indiana University, Bloomington, Indiana 47405-7102, USA.
[§] Institute of Physical and Theoretical Chemistry, University of Tübingen, 72076 Tübingen, Germany

**ABSTRACT:** High spin ($S$ = 3/2) organic triradicals may offer enhanced properties with respect to several emerging technologies, but those synthesized to date typically exhibit small doublet quartet energy gaps and/or possess limited thermal stability and processability. We report a quartet ground state triradical **3**, synthesized by a Pd(0)-catalyzed radical-radical cross-coupling reaction, which possesses two doublet-quartet energy gaps, $\Delta E_{DQ} \approx 0.2 - 0.3$ kcal mol$^{-1}$ and $\Delta E_{DQ}2 \approx 1.2 - 1.8$ kcal mol$^{-1}$. The triradical has a 70+% population of the quartet ground state at room temperature, and good thermal stability with onset of decomposition at >160 °C under inert atmosphere. Magnetic properties of **3** are characterized by SQUID magnetometry in polystyrene glass and by quantitative EPR spectroscopy. Triradical **3** is evaporated under ultra-high vacuum to form thin films of intact triradicals on silicon substrate, as confirmed by high resolution X-ray photoelectron spectroscopy. AFM and SEM images of the ~1-nm thick films indicate that the triradical molecules form islands on the substrate. The films are stable under ultra-high vacuum for at least 17 h but show onset of decomposition after 4 h at ambient conditions. The drop-cast films are less prone to degradation in air and have longer lifetime.

## INTRODUCTION

Organic radicals with high-spin ground states and large energy gap between the high-spin ground state and low-spin excited state are promising building blocks for organic magnets,[1-10] spintronics,[11] spin filters,[12-14] sensors,[15] memory devices,[16-19] and exploration of quantum interference on molecular conductance.[20] Their potential use in organic electronics depends not only on intrinsic electronic properties and stability but also processability. Albeit the design principles for high-spin radicals are clearly established[21,22] and a few triplet ground state diradicals with robust stability are prepared,[23-27] there are only few reports of isolated high-spin triradicals that process both good thermal stability and an energy gap between the high-spin ground state and low-spin excited state on the order of the thermal energy ($RT$) at room temperature.[15,28-30]

Recently, we reported the robust triplet ground state diradicals **1** and **2** (Figure 1).[31,32] To our knowledge, these and the analogous oxoverdazyl-based diradicals[33] are the only neutral high-spin diradicals that are well characterized by thermogravimetric analysis (TGA), to establish firmly their robust thermal stability.[34] Diradicals **1** and **2** have an onset of decomposition at 175 and 160 °C, while the onset at 192 °C was reported for the oxoverdazyl-based diradical.[31-33] Diradical **2**, which has a 95+% thermal population of the triplet state at room temperature, can be evaporated under ultra-high vacuum (UHV) to form thin films on silicon.[32]

The next challenge is a high-spin triradical, with a significant population of quartet ground state at room temperature, that is suitable for thin film fabrications. The design of triradicals involves another hurdle when an additional radical extends the molecular size. The triradical must not only possess excellent thermal stability but also a molecular mass that is still under the achievable limit of evaporation temperature.[35] From our experience,[32,35] the controlled evaporation of diradicals is already very demanding. It would be unprecedented to successfully fabricate the first thin-film of a high-spin triradical.[35-38]

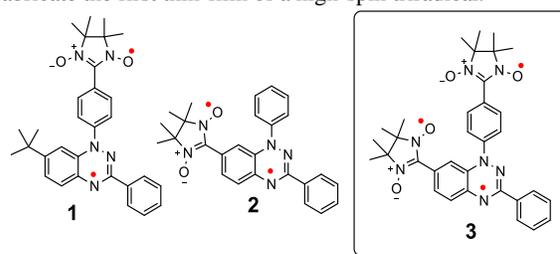

**Figure 1.** Thermally robust triplet ground state diradicals **1** and **2**, and quartet ground state triradical **3**.

Koutentis and co-workers recently reported the condensation-based synthesis of a Blatter-type triradical. The triradical is based on sequential connectivity of trimethylenemethane (TMM), $m$-phenylene, and TMM, thus it is expected to possess nearly degenerate quartet and doublet states, that is, three $S$ = ½ radicals are magnetically independent.[4] Their experimental finding of 300% spin concentration, corresponding to three $S$ = ½ spins, means that the triradical is composed of three $S$ = ½ spins that are magnetically independent at the temperature of measurement, in this case room temperature. Therefore, there is no experimental evidence for the claimed quartet ground state.[39]

The first pivotal step is the development of efficient synthetic methodologies for high-spin triradicals. Cross-couplings reactions are among the highly versatile and efficient methodology in organic synthesis, relying on a metal catalyst to form a bond between two different starting materials that usually are enriched with an activating group. The development of the cross-coupling reactions in the synthesis of organic radicals has faced tremendous obstacles, largely due to the inherent reactivity of organic radicals which profoundly affect both the activation of the starting materials and the search for a suitable catalyst.[40] Okada and coworkers recently demonstrated the Pd(0)-catalyzed cross-coupling between the gold(I)-(nitronyl nitroxide) complex, such as **4** (Scheme 1), with diamagnetic iodo-substituted aromatic compounds.[41,42] Tretyakov and coworkers also utilized complex **4** for the Pd(0)-catalyzed cross-coupling with iodo-oxoverdazyls, which led to oxoverdazyl-nitronyl nitroxide diradicals in high yields.[33] These results provide valuable insight and motivation to further develop the cross-coupling methodology for a variety of stable organic radicals. We focus on the cross-coupling reactions between Blatter radicals and nitronyl nitroxide radicals (Scheme 1).

The Blatter moiety is considerably more electron-rich, compared to typical diamagnetic π-systems or oxoverdazyl radicals (*vide infra*, Electrochemistry section). Thus, the common Pd(0) catalyst, such as Pd(PPh$_3$)$_4$ that was used both by Okada and Tretyakov,[33,41,42] may not be sufficiently powerful. In addition, spin delocalization in the Blatter radicals could be an impediment. The degree of spin density delocalization correlates with the strength of exchange coupling. The oxoverdazyl-nitronyl nitroxide diradicals have the smaller singlet-triplet energy gaps, $\Delta E_{ST}$ < 0.3 kcal mol$^{-1}$,[33] compared to $\Delta E_{ST} \approx 0.5$ kcal mol$^{-1}$ for **1** and $\Delta E_{ST} \approx 1.7$ kcal mol$^{-1}$ for **2**.[31,32] Thus, the spin density in the Blatter radicals is more delocalized than that in oxoverdazyls, leading to greater densities at the carbons of the C-I bonds in the starting Blatter radicals such as **5** and **6** (Scheme 1).

Here we report the synthesis and study of high spin ($S$ = 3/2) triradical **3** (Figure 1), in which we exploit the Pd(0)-catalyzed radical-radical cross-coupling reactions between di-iodo-substituted Blatter radical and nitronyl nitroxides. Triradical **3** has two doublet-quartet energy gaps, $\Delta E_{DQ} \approx 0.2 – 0.3$ kcal mol$^{-1}$ and $\Delta E_{DQ}2 \approx 1.2 – 1.8$ kcal mol$^{-1}$, i.e., same order of magnitude as the thermal energy at room temperature, thus possessing a quartet ground state that is 70+% populated at room temperature. Triradical **3** is thermally robust, with an onset of decomposition at ~160 °C under inert atmosphere and is thermally evaporated under ultra-high vacuum to form thin films on SiO$_2$/Si(111) wafers, with X-ray photoelectron spectroscopy indicating the presence of intact **3**. The AFM and SEM images of the evaporated films indicates the triradical molecules form isolated islands on the substrate, including preferential growth along a line defect of the substrate. We present here the preparation and characterization of the first thin film of high spin ($S$ = 3/2) organic triradical.

RESULTS AND DISCUSSION

**Synthesis of 3**. Syntheses of diradicals **1** and **2** start from the corresponding cyano-benzotriazinyl (cyano-Blatter) radicals, which are reduced to formyl-Blatter radicals, and then condensed with 2,3-bis(hydroxyamino)-2,3-dimethylbutane.[31,32,43,44] Oxidation of the condensation products yield diradicals **1** and **2**, with isolated yields of 4–12% and 18–29%, respectively, for the multistep syntheses.[31,32] An analogous approach to triradical **3**, starting from the corresponding di-cyano-Blatter radical produces only miniscule quantities of triradical with ~1% yield.

The cross-coupling of di-iodo-Blatter radical, such as **5**,[40,45-48] with 2+ equiv of **4** using the commonly used Pd(PPh$_3$)$_4$, with up to 60 mol% loading,[33,40-42] produces a low yield (6-17%) of triradical **3** while the coupling with 1+ equiv of **4** provides the mono-iodo-substituted diradical **8** in ~10% isolated yield (SI). Notably, the highly reactive Pd(0)-catalyst, Pd[(*t*-Bu)$_3$P]$_2$,[49] at a 30% mol loading in the cross-coupling reaction of **4** with **5** allows for synthesis of triradical **3** in one step in good isolated yields (Scheme 1). This approach enables a routine preparation of **3** in 100+ mg batches (SI).

We explore the cross-coupling reaction of **4** with mono-iodo-Blatter radical **6**[40,45-48] using the reactive catalyst, Pd[(*t*-Bu)$_3$P]$_2$,[49] which gives diradical **2** in 30 – 44% isolated yields (Scheme 1). We try the reactive Pd(II)-catalyst,[41] such as [1,3-bis(2,6-diisopropylphenyl)imidazol-2-ylidene](3-chloropyridyl)palladium(II) dichloride, commonly abbreviated as Pd-PEPPSI-*i*-Pr, in the reaction of **4** with **6**, but the reaction mixtures fail to yield any detectable diradical **2**.

We also examine the recently developed Pd(0)-catalyzed cross-coupling between diamagnetic iodo-substituted aromatics and nitronyl nitroxide **7** in the presence of a strong base (*t*-BuONa).[50] Starting from **6**, this methodology provides diradical **2** in a low yield (~10%). The cross-coupling of the di-iodo-Blatter radical **5** with **7** produces only mono-iodo-substituted diradical **8** in low yield (~1 – 10%), while triradical **3** is not detectable under various conditions (SI).

**Scheme 1. Synthesis of Triradical 3 and Diradicals 2 and 8.**

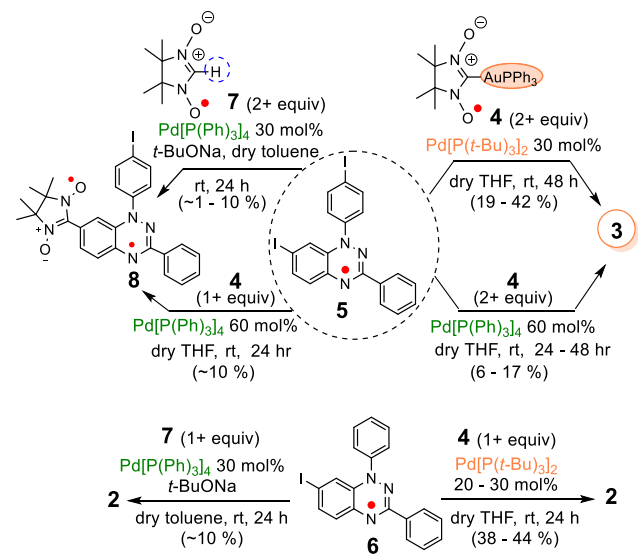

**X-ray crystallography.** The structure of triradical **3** is supported by single crystal X-ray analysis. Two pseudo polymorphs[51] are analyzed, triclinic (centrosymmetric P-1) and orthorhombic (Pna2$_1$), obtained by slow evaporation of solutions of **3** in toluene/heptane and chloroform/pentane, respectively. The orthorhombic structure is a solvent polymorph, containing one molecule of chloroform. The two pseudo polymorphs have significantly different crystal packing with the triclinic structure containing $C_i$-symmetric dimers of triradical molecules, while in the orthorhombic crystal structure, one-dimensional π-stacks of triradical molecules are formed (see: SI).[32-34,52]

In molecules of triradical **3**, both nitronyl nitroxide radical moieties are nearly coplanar with the 1,2,4-benzotriazinyl (Blatter) radical π-system (Figure 2); in both pseudo polymorphs, absolute values of N-C20-C5-C and N-C27-C11-C torsional angles are 19.8 – 22.5° and 28.6 – 34.1°, respectively. However, torsional angles between 1,4-phenylene rings (C8–C13) and the Blatter moiety are considerably greater, i.e., C9-C8-N1-C7 = 40.9 – 49.7° and C13-C8-N1-N2 = 49.3 – 51.4° (Figs. S2 and S5, SI). Consequently, the exchange coupling pathway from Blatter moiety to nitronyl nitroxide O1-N4-C20-N5-O2 (NNO1) is shorter and more co-planar than that to O3-N6-C27-N7-O4 (NNO2). Therefore, the exchange couplings associated with these two paths, $J_1/k$ and $J_2/k$, are expected to be significantly different (Figure 2).

**EPR spectroscopy.** The EPR spectra of **3** in glassy matrices show a quartet ($S = 3/2$) state, with a small admixture of thermally populated doublet ($S = \frac{1}{2}$) state at $T = 110$ K (Figure 3). Because of relatively small value of zero-field splitting parameter, $D \approx 80$ MHz (Table 1), only a weak half-field ($|\Delta m_S| = 2$) transition can be observed and no $|\Delta m_S| = 3$ signal can be detected.[53-55] Spectral simulations confirm the purity of the triradical,[56] that is, the absence of $S = 1$ or $S = 1/2$ impurities. The spectral width for $S = 3/2$ triradical **3** in toluene/chloroform glass is $4D \approx 320$ MHz, which is intermediate between $2D \approx 140$ MHz for $S = 1$ diradical **1** and $2D \approx 480$ MHz for **2**. This reflects the intermediate strength of magnetic dipole-dipole interactions in **3** that dominate the EPR spectra in glassy matrices. Similar to diradical **2**, the B3LYP/EPR-II calculations of **3** not only provide an overestimated value of $D = 160$ MHz but also indicate the positive sign of $D$, which is inconsistent with the experimental EPR spectrum (SI).[32,57-62]

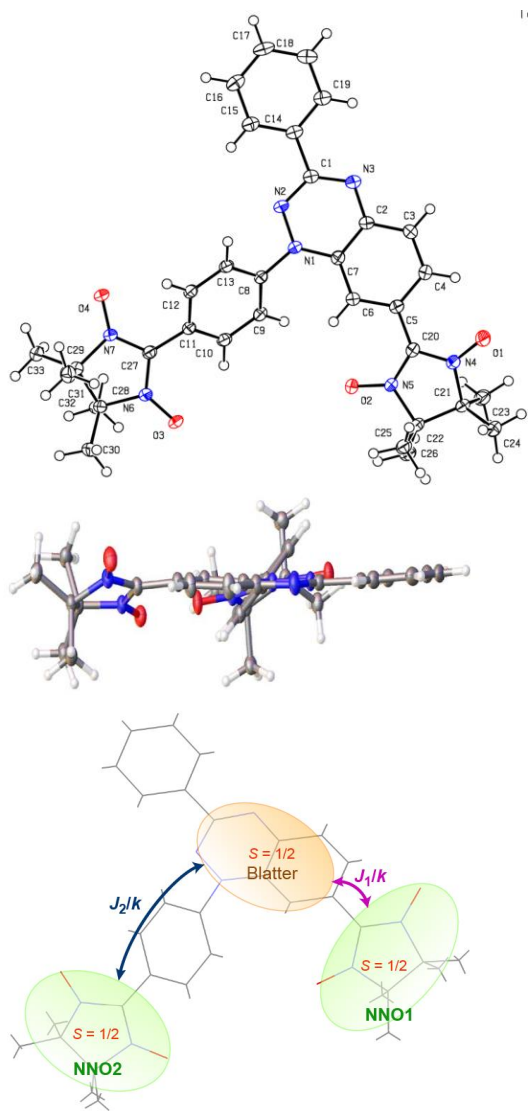

**Figure 2.** Single crystal X-ray structure of triradical **3** (triclinic, centrosymmetric P-1 polymorph). <u>Top</u>: Top view, Ortep plot with carbon, nitrogen, and oxygen atoms depicted using thermal ellipsoids set at the 50% probability level. <u>Middle</u>: Side view of **3**. <u>Bottom</u>: Exchange coupling in **3**. Additional details may be found in the SI; crystallographic data were deposited: CCDC #s: 2062663 and 2062664.

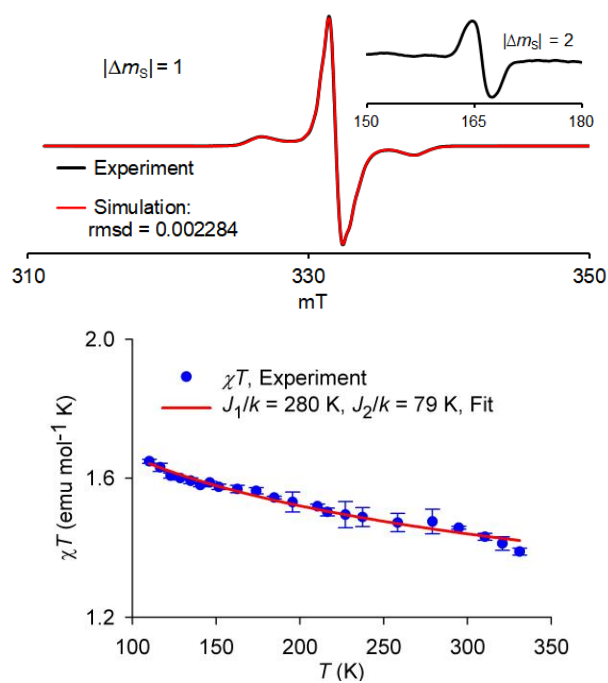

**Figure 3**. EPR spectroscopy of triradical **3**. <u>Top</u>: EPR ($\nu = 9.3245$ GHz) spectrum of 2.32 mM triradical **3** in 2-MeTHF glass at 105 K. The $|\Delta m_S| = 2$ transition is shown as an inset. Simulation of the $|\Delta m_S| = 1$ region: $S = 3/2$, weight = 1.0000, $D = 77.84$ MHz, $E = 22.70$ MHz, $g_{xx} = 2.0075$, $g_{yy} = 2.0047$, $g_{zz} = 2.0065$; $H$-strain (MHz): $H_x = 15.2$, $H_y = 79.3$, $H_z = 47.2$; $S = 1/2$, weight = 0.17245, $g_{xx} = 2.0018$, $g_{yy} = 2.0137$, $g_{zz} = 1.9968$; $H$-strain (MHz): $H_x = 10.2$, $H_y = 7.1$, $H_z = 17.8$. <u>Bottom</u>: Quantitative EPR spectroscopy of 1.36 mM **3** in toluene/chloroform (3:1). Experimental values of $\chi T$ (mean ± SE, $n = 3$) in the $T = 110 – 331$ K range and numerical two-parameter fit with two variable parameters, $J_1/k = 280 \pm 17$ K and $J_2/k = 79 \pm 2.7$ K (mean ± SE). Further details are reported in the SI: Eq. S1 and Figs. S12–S18.

We carry out variable temperature quantitative EPR spectroscopy on **3** in toluene/chloroform, 3:1.[31,63] At each temperature in the $T = 110 – 331$ K range, three independent measurements of the sample and the spin counting reference (Tempone in toluene/chloroform, 3:1) are obtained. The resultant average values of $\chi T$ (mean ± SE, $n = 3$) are fit to the non-symmetrical triradical model (Figure 2, eq. S1, SI)[30,64] using two variable parameters, exchange coupling constants, $J_1/k$ and $J_2/k$ and one fixed parameter, weight correction factor, $N = 0.99$ (Figure 3). The parameter $N$ accounts for an inaccurate concentration of **3**, due to the uncertainty in weighing about 1 mg of triradical sample. The value of $N$ is derived from the spectral simulations (fits) of EPR spectra at 110 K. These fits provide the relative content of $S = 3/2$ ground state and $S = ½$ excited state, which enables an estimation of the value of $\chi T$. The ratio of the measured $\chi T = 1.632 \pm 0.0058$ emu K mol$^{-1}$ ($n = 3$) to the estimated $\chi T = 1.6479 \pm 0.0003$ emu K mol$^{-1}$ ($n = 3$) provides the value of $N = 0.99$ (Figs. S18 and S19, SI).

Values of $J_1/k = 280 \pm 17$ K and $J_2/k = 79 \pm 2.7$ K (mean $\pm$ SE) (Figure 4), obtained from a numerical fit to the non-symmetrical trimer model (Figure 2 and eq. S1, SI), allow for the calculation of doublet-quartet energy gaps, $\Delta E_{DQ} \approx 0.2$ kcal mol$^{-1}$, for the lowest $S = ½$ excited state and $\Delta E_{DQ}2 \approx 1.2$ kcal mol$^{-1}$, the second lowest $S = ½$ excited state, using Eqs. 1 and 2,[30] respectively (Table 1).

$$\Delta E_{DQ} = J_1 + J_2 - [J_1^2 + J_2^2 - (J_1 * J_2)]^{1/2} \quad (1)$$
$$\Delta E_{DQ}2 = J_1 + J_2 + [J_1^2 + J_2^2 - (J_1 * J_2)]^{1/2} \quad (2)$$

**Table 1. Magnetic characterization of triradical 3 vs diradicals 1 and 2.**

| | | matrix | $|D|$ (MHz) | $|E|$ (MHz) | $J_1/k$ (K) | $J_2/k$ (K) | $\Delta E_{DQ}{}^a/\Delta E_{ST}$ (kcal mol$^{-1}$) |
|---|---|---|---|---|---|---|---|
| **3** | EPR | Tol/Chl[b] | 80 | 23 | 280 | 79 | +0.22[a] |
| | EPR | 2-MeTHF | 78 | 23 | - | - | - |
| | EPR | Polystyrene | 77 | 23 | - | - | - |
| | SQUID | Polystyrene | - | - | 425 | 109 | +0.30[a] |
| | DFT | Gas phase | 160[c] | 20[c] | - | - | +1.14[d] |
| **2** | SQUID | Crystals[32] | - | - | 438 | - | +1.74 ± 0.07 |
| | SQUID | Polystyrene[32] | - | - | 419 | - | +1.68 ± 0.16 |
| | EPR | Tol/Chl[b,32] | 242 | 35.1 | - | - | - |
| **1** | SQUID | Crystals[32] | - | - | - | 126 | +0.50 ± 0.02 |
| | EPR | Tol/Chl[b,31] | 69.6 | 4.2 | - | 117 | +0.47 |

[a] $\Delta E_{DQ}$ (kcal mol$^{-1}$) determined experimentally using eq 1. [b] Toluene/chloroform, 3:1. [c] $D$ and $E$ computed at the B3LYP/EPR-II level using ORCA.[57] [d] BS-DFT-computed $\Delta E_{DQ}$ at the UB3LYP/6-31G(d,p)+ZPVE level;[65] $\Delta E_{DQ}2 = 1.22$ (EPR) and 1.83 (SQUID) kcal mol$^{-1}$ vs. 2.59 (DFT) kcal mol$^{-1}$.

**SQUID magnetometry.** The quartet ground state of **3** is confirmed by SQUID studies. The $\chi T$ vs $T$ and $M/M_{sat}$ vs $H/(T - \theta)$ plots for a 19 mM sample of **3** in polystyrene are shown in Figure 4.

Using the non-symmetrical triradical model, which includes the effects of paramagnetic saturation (Figure 2, eq S1, SI), the fits to the $\chi T$ vs. $T$ data in the $T = 1.8 - 370$ K range give the following values of variable parameters: $N = 0.718 \pm 0.002$, $J_1/k = 425 \pm 37$ K and $J_2/k = 109 \pm 4$ K (mean $\pm$ SE), where $N$ corresponds to weight correction factor (Figure 4A). The values of both $J_1/k$ and $J_2/k$ are greater than those obtained by EPR spectroscopy in toluene/chloroform. Consequently, larger $\Delta E_{DQ} \approx 0.30$ and $\Delta E_{DQ}2 \approx 1.83$ kcal mol$^{-1}$ are obtained using Eqs. 1 and 2 (Table 1). Also, these fits indicate that the intermolecular exchange interactions between molecules of **3** are negligible, i.e., the value of mean-field parameter, $\theta$, is near zero.

Using Brillouin functions with a small mean-field parameter, $\theta = -0.06$ K, two-parameter fits to the magnetization ($M$) vs. magnetic field ($H$) data, i.e., the $M$ vs. $H/(T - \theta)$ data, at low temperatures ($T = 1.8 - 5$ K) provide the total spin, $S = 1.5$ and magnetization at saturation, $M_{sat} = 0.736$ $\mu_B$ ($\mu_B$ = Bohr magneton). These data unequivocally confirm a quartet ($S = 3/2$) ground state for **3**. In addition, the value of $M_{sat} = 0.74$ $\mu_B$ is comparable to $N =$ 0.72, obtained from the fit to the $\chi T$ vs. $T$ data, thus indicating that the triradical is pure. Both $M_{sat}$ and $N$ are less than 1.00 because of weighing errors of **3** in the sub-milligram range.

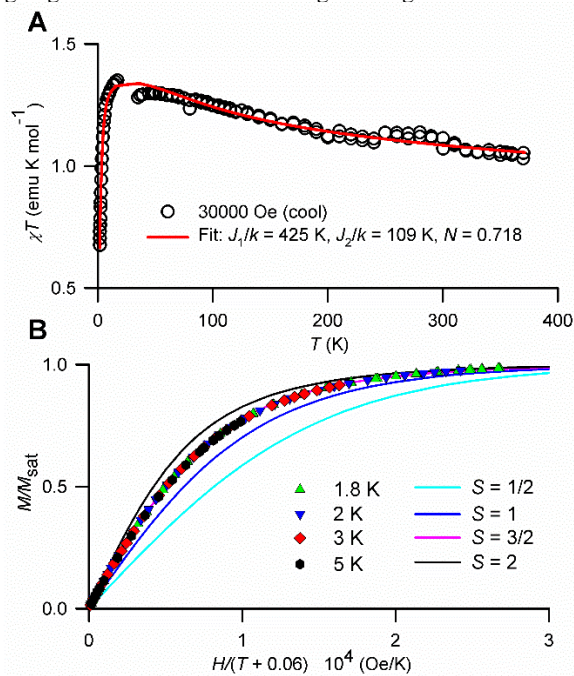

**Figure 4.** SQUID magnetometry of 19 mM triradical **3** in polystyrene matrix. **A**: $\chi T$ vs $T$ data at $H = 30000$ Oe in the cooling mode were fit to a triradical model (eq S1, SI), using three variable parameters: $N = 0.718 \pm 0.002$, $J_1/k = 425 \pm 37$ K and $J_2/k = 109 \pm 4$ K (mean $\pm$ SE), where $N$ corresponds to weight correction factor. Downward turn in the $\chi T$ vs $T$ plot at low $T$ is predominantly due to paramagnetic saturation. The gap in the data around 25 K corresponds to the passage through $M = 0$ for the entire sample. **B**: $M/M_{sat}$ vs $H/(T - \theta)$ plot, where $\theta = -0.06$ K, at $T = 1.8-5$ K (symbols) and the Brillouin curves corresponding to $S = 1/2-2$ (lines). Further details are reported in the SI: Figs. S20 and S21.

A sample of **3** in benzene is investigated, in which the fits to $M$ vs. $H/(T - \theta)$ data provide the total spin, $S = 1.5$ and indicate an $S = 3/2$ ground state (Fig. S23, SI). Due to the relatively large mean-field parameter, $\theta \approx -1$ K, indicating relatively strong intermolecular antiferromagnetic interactions, the 4-parameter fits to the $\chi T$ vs. $T$ data in the relatively narrow $T = 1.8 - 260$ K range are unreliable, and thus, $\Delta E_{DQ}$ could not be determined.

**Electrochemistry and UV-vis-NIR spectroscopy.** Cyclic voltammetry for diradical **2** and triradical **3** in 0.1 M tetrabutylammonium hexafluorophosphate in dichloromethane at room temperature shows the presence of reversible waves associated with the oxidation of Blatter radical moieties ($E^{+/0} \approx +0.4$ V) and nitronyl nitroxide moieties ($E^{2+/+} \approx +1.0$ V). A reversible wave corresponding to the reduction of either the Blatter radical or the nitronyl nitroxide moiety is also observed in the $E^{-/0} \approx -0.7 - (-0.9)$ V range (Table S5 and Figs. S9–S11). (All redox potentials are reported versus SCE.) These values are comparable to those obtained for the parent Blatter radical (+0.10 and –0.96 V) and a derivative of nitronyl nitroxide (+0.81 and -0.75 V) in acetonitrile.[18,66] Notably, $E^{+/0}$ for the oxoverdazyl radicals is about 0.4 V more positive,[33,67] and therefore they are considerably more difficult to oxidize.

UV-vis-NIR spectra of diradical **2** and triradical **3** in dichloromethane have a similar spectral pattern, consisting three major bands at 300, 370 – 380, and 510 – 550 nm (Fig. S8, SI). The peak intensity at 371 nm ($\varepsilon_{max} = 2.26 \times 10^4$ L mol$^{-1}$ cm$^{-1}$) for **3** is about twice of that at 381 nm ($\varepsilon_{max} = 1.3 \times 10^4$ L mol$^{-1}$ cm$^{-1}$) for diradical **2**. The intense 371 nm peak is associated with the band originating from the phenyl-substituted nitronyl nitroxide, which usually appears at 362 nm ($\varepsilon_{max} = 1.77 \times 10^4$ L mol$^{-1}$ cm$^{-1}$) in hexane or at 360 nm ($\varepsilon_{max} = 1.33 \times 10^4$ L mol$^{-1}$ cm$^{-1}$) in ethanol.[43,68] Diradical **2** and triradical **3** have nearly identical absorption onsets in the 860–880 nm range, corresponding to an optical gap, $E_g = 1.42 \pm 0.01$ eV (mean ± SE). The UV-vis-NIR spectrum for triradical **3** could be reproduced by the TD-DFT computation at the UCAM-B3LYP/6-31+G(d,p)/IEF-PCM-UFF level of theory employing a dichloromethane solvent model (Fig. S26, SI).

**Thermal stability.** Triradical **3** possesses excellent stability at ambient conditions. It can be purified by chromatography using normal phase silica gel. Thermogravimetric analysis data suggest that thermal decomposition of **3** starts at 166 °C (1% mass loss), which is slightly higher than the onset temperature for diradical **2** (Figure 5) and lower than that for diradical **1**. The maximum rate of mass loss for **3** is at 180 °C.

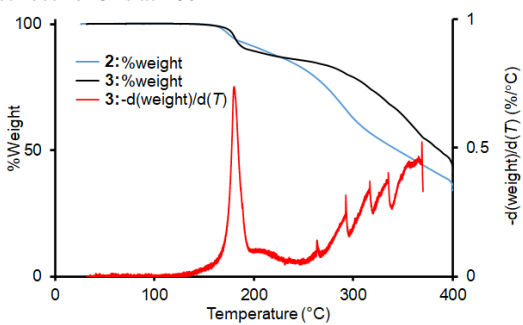

**Figure 5.** Thermogravimetric analysis (TGA) of triradical **3** under N$_2$; heating rate = 5 °C min$^{-1}$.

**Thin films of 3 on SiO$_2$/Si(111) substrate.** An important prerequisite of the technological applications foreseen for this class of materials requires attaching molecules to a substrate, either to form an interface or a film. A controlled and clean way to achieve this goal is using controlled evaporation. We have previously focused our efforts on the evaporation of radicals and diradicals.[32, 35-38,69-72]

Here, we extend our protocol to evaporate triradical **3**. We deposit thin films of triradical **3** on SiO$_2$/Si(111) wafers by organic molecular beam deposition (OMBD),[73] which has been proven to be a suitable method for growing radical and diradical thin films. However, increasing the number of radical sites implies an increased reactivity during evaporation making it extremely challenging.[35,38] We investigate the obtained thin films by X-ray photoelectron spectroscopy (XPS), an effective and powerful tool for studies of organic and organic radical thin films.[38] We adopt the approach that was previously used for the diradical thin films to assess the intactness of triradical **3** in the films.[32,35] We also obtain the films by drop-casting deposition. Fabrication of films by the two growth methods allows exploring the differences due to preparation and thickness range. The C 1s and N 1s core level curves are shown in Figure 6 that also includes the fit components. In fact, XPS is sensitive to the stoichiometry of the films; identifying the contributions associated to each element in its chemical environment helps to gain information on the chemical composition of the films after evaporation (The fit procedure is described step by step in refs.[32,74]). The elemental concentration of the films calculated from the XPS spectra agrees very well with the stoichiometry of the triradical (carbon 84.4 % and nitrogen 16.6% as obtained from XPS versus the stoichiometric 82.5 % and 17.5 %, respectively). This agreement is further supported by the fit results of the single contributions due to photoelectrons emitted by atoms with different chemical environment that shows the expected intensities in the main line (Tables S6 and S7, SI). Note that we do not consider the O 1s spectroscopic line because it is a superposition of the signal from the films and the substrate making its fit analysis speculative.

The C 1s spectrum of the evaporated films is characterized by a main line at around 285.5 eV due to photoelectrons emitted from the atoms in the aromatic ring and the carbon atoms bound to hydrogen atoms (C-C, C-H and CH$_3$). The shoulder at higher binding energy is due to contributions from the electrons emitted from carbon atoms bound also to nitrogen (C-N).

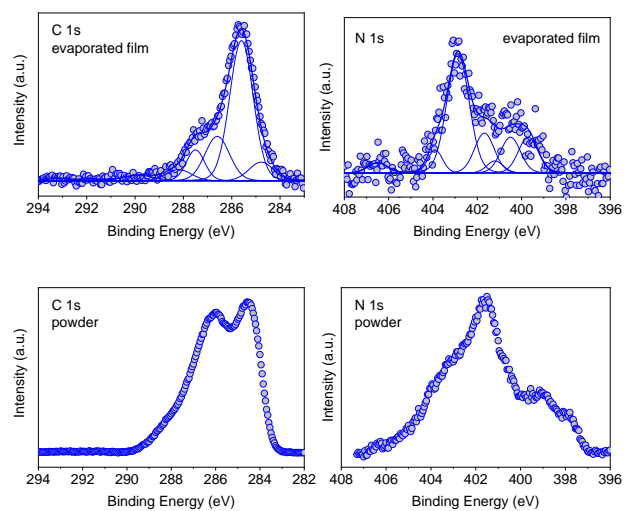

**Figure 6.** Typical C 1s and N 1s core level XPS spectra of triradical **3** deposited on SiO$_2$/Si(111) wafers (0.4 nm thick film, nominal thickness) (top plots) together with the fit analysis, compared to the powder spectra (bottom plots).

Nitrogen atoms, because of their higher electronegativity, shift the electronic cloud, thus, the electrons are emitted with lower kinetic energy, i.e., higher binding energy. The N 1s core level spectrum shows contributions due to seven nitrogen atoms: the three nitrogen atoms belonging to the Blatter radical have different chemical environment, while the two nitrogen atoms belonging to the nitronyl nitroxide (NN) radical have an equivalent chemical environment.[32,35,72,75] These differences give rise to a complex spectrum with two broad features, showing the highest intensity at around 402.5 eV that corresponds to the line expected in the NN radical N 1s core level spectrum.[72] We also observe the presence of satellites, they are typical features in photoemission that appear as an effect of the relaxation processes due to the creation of a core-hole.[76,77] Based on the comparison of the film fit results (Tables S6 and S7) and the molecular stoichiometry, we can conclude that there is no degradation of the triradical molecules during evaporation and deposition, under the present conditions.

The comparison with the powder spectra also indicates that the evaporation and the deposition of the film are successful. We note that the XPS curves of the powder are affected by a strong charging effect. This effect is expected in organic crystals because of the absence of efficient screening of the core-hole.[78] Here this effect is very strong leading to a broadening of the features and changes in the line intensities. We noticed similarly strong charging effects also in the XPS curves of diradical **2** that may be viewed as a fusion of the Blatter radical with a single nitronyl nitroxide radical, therefore, having two radical moieties in common with triradical **3**.[32]

XPS also offers the opportunity to identify *in-situ* the thin film growth mode following the decay of the substate signal during evaporation. We follow the XPS core level signal of the substrate (Si 2p) by looking at its attenuation upon film deposition (Figure 7). The curve is characterised by a very slow decay. This intensity trend hints at a Volmer-Weber (VW) growth mode, i.e., island growth.[79] This result is consistent with the *ex-situ* atomic force microscopy (AFM) and the scanning electron microscope (SEM) images obtained for triradical **3** films (Figure 7), which are clearly showing a film morphology dominated by islands. The VW growth mode occurs when the interaction between the deposited molecules is much stronger than between the molecules and the substrate.[80] We have observed this growth mode for all thin films of radicals and diradicals that we have previously investigated, grown on $SiO_2$/Si(111) wafers, keeping the substrate at room temperature. However, the tendency to grow isolated islands increases from radicals to diradicals,[32,35] and now it is confirmed by the present results on triradical **3**.

triradical. The fit procedure supports the expected result of films having elemental concentrations in agreement with the stoichiometry of the triradical (Tables S8 and S9, SI). The AFM images as well as the SEM images are featureless and flat (over the examined field of view) as expected for the drop-cast preparation. The SEM images that comprise a larger area are characterized by distributed circular valleys, as typically observed in films obtained using this processing method, due to the drying effects caused by the solvent evaporation (Figure 8).

Finally, we monitor the lifetime of the films in UHV (base pressure $2 \times 10^{-10}$ mbar) by using XPS, focusing on the N 1s core level spectrum that is correlated with the nitronyl nitroxide and Blatter radicals.[38,72] We adopt the protocol previously applied to radical and diradical films.[32,35,74] We observe no major changes in the spectra of the evaporated films after their exposure to UHV at room temperature for around 17 h (Fig. S25, SI). However, after 4 h of air exposure, we observe major changes in their XPS spectra, indicating film degradation (Fig. S25, SI). Thus, the triradical thin films are much less robust in air. We have found that diradicals thin films have shorter lifetime in air than their monoradical analogues (nitronyl nitroxide and the Blatter monoradical derivatives showed changes in their XPS after the films were kept for several weeks/months at ambient conditions [35,37,72]). This tendency is now also found for the evaporated triradical **3** films.

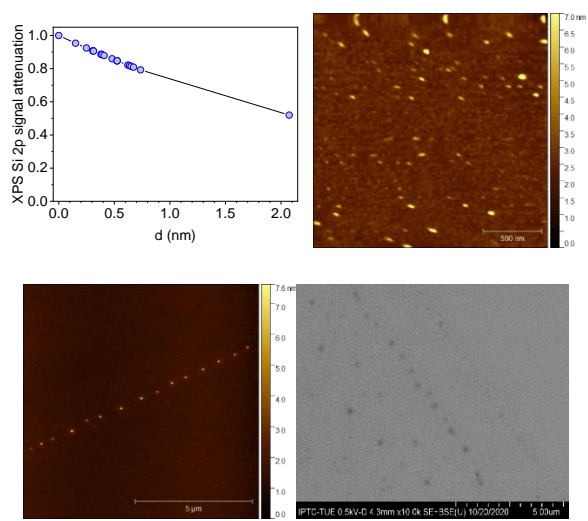

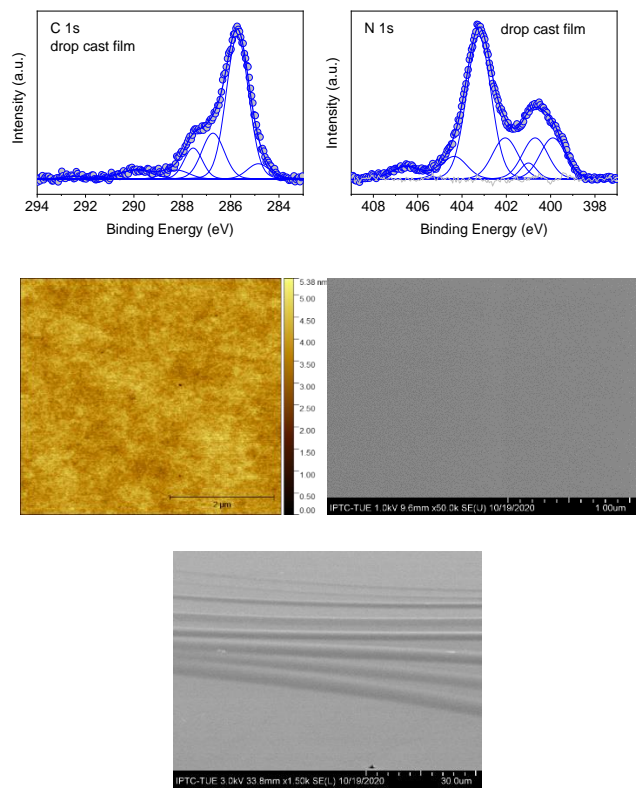

**Figure 7.** (Upper panels) Attenuation of the Si 2p XPS signal, normalized to the corresponding saturation signal, as a function of film nominal thickness, deposition at room temperature (the line is a guide to the eye). A typical 2 μm x 2 μm AFM image of a film (0.4-nm nominally thick film, bottom panel). (Lower panels) A 10 μm x 10 μm AFM image of a film (0.7-nm nominally thick film,) showing the island decorating a line defect of the substrate, together with the corresponding SEM image.

We also observe that the non-stoichiometric films, resulting from failed evaporations of **3**, show a different island morphology. The microscopy investigations (Fig. S24, SI) suggest that different stoichiometry leads to different molecule-molecule interactions and plays a central role in the film morphology.

We investigate the drop-cast films using the same techniques. XPS spectra of the film (Figure 8) show the same features as those in the spectra of the evaporated samples. The results further affirm that the evaporation does not degrade the

**Figure 8**. (Upper panels) C 1s and N 1s core level XPS spectra of a drop-cast film of triradical **3** deposited on $SiO_2$/Si(111) wafers (top plots) together with the fit analysis. (Middle panels) A typical featureless 5 μm x 5 μm AFM image of a drop-cast film with the corresponding SEM image. (Lower panel) A typical SEM image showing the circular valleys, due to the drying effects.

The drop cast film N 1s core level spectra change on a longer time scale, when exposed to air. These changes are different,

however, rather than showing different line shapes as in the case of the evaporated films, there is a change in the relative intensities of the two broad features (Fig. S25, SI), which may be due to structural changes. We also observe changes in the intensity of the satellite features that would support this hypothesis.[81,82] In addition, the changes in N 1s core level spectra tend to saturate after 84 hours. Also, the adsorption of ambient nitrogen may contribute to the changes of the core level line intensities. Because the drop cast films are thicker by two orders of magnitude than the evaporated films (nm versus hundreds of nm), we may speculate that the changes in intensities originate from the degradation of the more superficial layers of the drop cast films. These degraded layers then act as a protective buffer for the underneath material that keeps its properties.

CONCLUSION

We have synthesized a high-spin triradical **3** by the radical-radical cross-coupling reaction between di-iodo-substituted Blatter radical and nitronyl nitroxide derivative. We show that at room temperature, 70+% of the triradical molecules populate the high-spin, $S = 3/2$, ground state. The triradical possesses a remarkable thermal stability to permit fabrication of intact triradical thin films on silicon substrate via evaporation under ultra-high vacuum. The triradical molecules form isolated islands on the substrate with the tendency to decorate the substrate line defects, which might be useful for fabrication of nanostructured functionalized surfaces. The triradical films are stable under ultra-high vacuum, however, within few hours of exposure to air, XPS of the films show major changes. The drop-cast films show longer air lifetime. We have demonstrated that it is possible to evaporate triradicals and deposit their thin films under controlled conditions without degradation. The triradical films are less stable compared to the films of diradical **2** or nitronyl nitroxide and Blatter monoradicals. Our triradical, with an unprecedented combination of high-spin ground state and thermal properties, that is suitable for thin film fabrication under ultra-high vacuum, could facilitate the development of purely organic magnetic and electronic materials.

EXPERIMENTAL SECTION

(Nitronyl nitroxide-2-ido)(triphenylphosphine) gold(I) **4** is synthesized by the reaction of nitronyl nitroxide **7** and Au$^I$(PPh$_3$)Cl with NaOH in methanol/dichloromethane.[41,83,84] After purification by chromatography (deactivated Al$_2$O$_3$), the spin concentration of **4** is up to 99%. Di-iodo-Blatter radical **5** and its mono-iodo analogue **6** are prepared according to previous literature.[45,46]

Frozen and liquid solution EPR spectra were obtained using a Bruker EMX-plus X-band spectrometer and simulated with the EasySpin software.[56] The TGA instrument (TA Instruments TGA 550) was run either without or with IR attachment (Thermo NICOLET Is50 NIR). Variable temperature (from 1.8 K to up to 370 K) magnetic susceptibility measurements of **3** were performed using a Quantum Design SQUID magnetometer with applied magnetic fields of 30 000 and 5000 Oe. Variable field (0 – 50,000 Oe) magnetization studies were carried out at temperatures of 1.8 – 5 K. Sample tubes for SQUID studies in dilute matrices[85] are described in the SI.

**X-ray crystallography.** Crystals of **3** for X-ray studies were prepared by slow evaporation from solution in in toluene/heptane and chloroform/pentane. Data collections were performed at 100 K at either the Advanced Photon Source, Argonne National Laboratory using $\lambda$ = 0.41328 Å synchrotron radiation (silicon monochromators) or the Indiana University using Mo Kα radiation. Following integration (SAINT),[86] the intensity data were corrected for absorption (SADABS).[87] The space groups for two polymorphs, P-1 and Pna2$_1$, were determined based on intensity statistics and the systematic absences. The structures were solved (SHELTX)[88] and then refined on $F$2 (SHELXL).[89] Crystal and structure refinement data for two polymorphs of **3** are in the Supporting Information and the deposited (CCDC #s: 2062663 and 2062664) files in CIF format.

**Synthesis of triradical 3.** Standard techniques for synthesis under inert atmosphere (argon or nitrogen), using custom-made Schlenk glassware, custom-made double manifold high vacuum lines, argon-filled MBraun glovebox, and nitrogen-filled glovebags. Chromatographic separations were carried out using either normal phase silica gel or neutral alumina.

Triradical **3**. A mixture of Blatter radical **5** (200.0 mg, 0.373 mmol, 1 equiv) and nitronyl nitroxide radical **4** (500.0 mg, 0.821 mmol, 2.2 equiv) was added to a Schlenk tube and evacuated on vacuum line for several minutes. The tube was kept under vacuum, and then it was transferred to the antichamber of the glove box. Inside the argon-filled glove box, palladium(0) catalyst (Pd[P(t-Bu)$_3$]$_2$, 57.0 mg, 30 mol%) was added into the tube. Freshly distilled, dry THF was added into the tube under an argon gas flow. The reaction mixture was stirred at room temperature for 48 hours. The dark red solution was evaporated, to provide a dark solid, which was purified by silica gel column chromatography using dichloromethane, followed by 10% ethyl acetate in dichloromethane, as eluents. The resultant solid was washed with pentane to give 93.1 mg (42%) of product **3** as a dark solid. TLC (silica gel, ethyl acetate): $R_f$ 0.43. HR-TOF-MS: $m/z$, [M + Na]$^+$ calcd for C$_{33}$H$_{36}$N$_7$O$_4$Na 617.2726; found, 617.2728 (0.3 ppm, RA = 100%). IR (powder, cm$^{-1}$): 2981.51, 2937.11, 1599.50, 1518.83, 1480.79, 1448.57, 1419.16, 1392.11, 1363.07, 1310.84, 1271.29, 1254.60, 1194.33, 1130.93, 1069.11, 1025. 05, 1017.26, 952.98, 919.95, 863.37, 834.99, 823.22, 788.48, 699.98. 2981.51, 2937.11, 1599.50, 1518.83, 1480.79, 1448.57, 1419.16, 1392.11, 1363.07, 1310.84, 1271.29, 1254.60, 1194.33, 1130.93, 1069.11, 1025. 05, 1017.26, 952.98, 919.95, 863.37, 834.99, 823.22, 788.48, 699.98. The spin concentration was determined to be 378% ($\chi T$ = 1.42 emu K mol$^{-1}$) in toluene/chloroform (4:1) fluid solution at 294 K and 435% ($\chi T$ = 1.63 emu K mol$^{-1}$) at 110 K in toluene/chloroform (3:1) glass. For each measurement, TEMPONE in the identical solvent was used as a spin counting reference. Because we use $S = ½$ monoradical as reference (100% spin concentration), for an $S = 3/2$ triradical with perfectly populated quartet ground state spin concentration should be 100%[3/2*(3/2 + 1)]/[1/2*(1/2 + 1)] = 500%; in other words, the signal intensity (or $\chi T$) should correspond to 5 independent monoradicals. Values of <500% reflect thermal population of the excited doublet states.

**Computational details.** All geometry optimizations for **3** were carried out at the UB3LYP/6-31G(d,p) level of theory, in the gas phase or with Gaussian 16 default IEF-PCM solvent model for toluene, tetrahydrofuran, or ethanol (Table S10, SI). Obtained minima were confirmed by frequency calculations. The broken-symmetry approach was applied for open-shell doublet calculations and spin contamination errors were corrected by approximate spin-projection method.[90,91] Broken symmetry doublet wavefunctions ($<S^2> \approx 1.8$) at the UB3LYP/6-31G(d,p)+ZPVE level of theory were checked for stability. All calculations were performed with the Gaussian 16 program suite.[65]

**Thin film growth, XPS and microscopy measurements.** Thin films were deposited on native $SiO_2$ grown on single-side polished n-Si(111) wafers. The films were evaporated via organic molecular beam deposition (OMBD) using a Knudsen Cell under UHV conditions. The UHV apparatus used for deposition and XPS measurements is composed of dedicated preparation and measurement (base pressure $3 \times 10^{-10}$ mbar) chambers. The measurement chamber is equipped with a monochromatic Al Ka source (SPECS Focus 500) and a hemispherical electron analyser (SPECS Phoibos 150). The wafers were cleaned by sonication for an hour each in ethanol and acetone and subsequent annealing at ca. 500 K in UHV and verification of cleanness by XPS. The nominal films thicknesses of the evaporated films were calculated from the attenuation of the substrate signal. Drop-cast films of triradical **3** from 50 μL of toluene solution with a concentration of 2-3 mg/mL were deposited on the substrates with an area of about 1 cm$^2$ under ambient conditions. Their thickness (100-250 nm) was estimated with a scratch-test using SEM images. For the preparation of powder samples indium foil was used into which the powder was firmly pressed to minimise the charging of the crystals. Pass energies of 50 eV and 20 eV were used for survey spectra and individual core level spectra, respectively. Spectra were calibrated to the Si 2p signal at 99.8 eV and the In 3d signal at 103.3 eV depending on the used substrate. Radiation damage was minimised by only measuring freshly prepared samples and limiting beam exposure. Beam exposure in measurements aimed to investigate the stability of the films was further reduced to attribute changes solely to the degradation by air or in UHV. This leads to a worse signal to noise ratio. Atomic force microscopy (AFM) was measured in air with a Digital Instruments Nanoscope III Multimode AFM using tapping mode and ScanAsyst® mode. Scanning electron microscopy (SEM) was measured using a HITACHI SU8030 ultra-high resolution field emission scanning electron microscope.

## ASSOCIATED CONTENT

### Supporting Information

General procedures and materials, additional experimental details, X-ray crystallographic files for **3** in CIF format, fit results for the energy positions and relative intensities of the photoemission lines in the C 1s and N 1s spectra. AFM and SEM microscopy of thin films. UHV and air film lifetime. This material is available free of charge via the Internet at http://pubs.acs.org

## AUTHOR INFORMATION

**\* Corresponding Authors**
arajca1@unl.edu
benedetta.casu@uni-tuebingen.de

Notes
The authors declare no competing financial interests.

## ACKNOWLEDGMENT

We thank the National Science Foundation (NSF), Chemistry Division for support of this research under Grants No. CHE-1665256 and CHE-1955349 (AR), and the National Institutes of Health (NIGMS #R01GM124310-01 to S.R. and A.R.) for the upgrade of EPR spectrometer. Support for the acquisition of the Bruker Venture D8 diffractometer through the Major Scientific Research Equipment Fund from the President of Indiana University and the Office of the Vice President for Research is gratefully acknowledged. NSF's ChemMatCARS Sector 15 is supported by the Divisions of Chemistry (CHE) and Materials Research (DMR), National Science Foundation, under grant number NSF/CHE-1834750. Use of the Advanced Photon Source, an Office of Science User Facility operated for the U.S. Department of Energy (DOE) Office of Science by Argonne National Laboratory, was supported by the U.S. DOE under Contract No. DE-AC02-06CH11357. We also would like to thank Thomas Chassé for accessing the photoelectron laboratory at the University of Tübingen. Financial support from the German Research Foundation (DFG) under the contract CA852/11-1 is gratefully acknowledged. We thank Dr. N. M. Gallagher for initial trials in the synthesis of triradical **3** using condensation method and Dr. Hui Zhang for sample preparations for SQUID measurements.

Insert Table of Contents artwork here

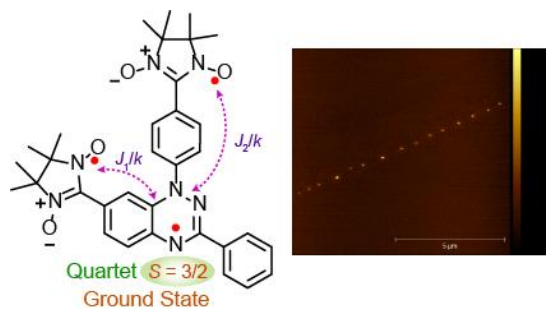